\documentclass{ws-p10x7}

\begin{document}

\newcommand{\nn}{\nonumber}
\newcommand{\fr}{\frac}
\newcommand{\gp}{p\!\!\!/}
\newcommand{\gk}{k\!\!\!/}
\newcommand{\gm}{m\!\!\!\!/}
\newcommand{\eq}[1]{(\ref{#1})}
\newcommand{\ab}{( \alpha - \beta )} 
\newcommand{\mh}{m_h^2}
\newcommand{\mH}{m_H^2}
\newcommand{\mg}{m_G^2}
\newcommand{\ma}{m_A^2}
\newcommand{\mt}{m_t^2}
\newcommand{\mw}{m_W^2}
\newcommand{\mz}{m_Z^2}
\newcommand{\vt}{v^3}
\newcommand{\p}{p^2}
\newcommand{\al}{\alpha}
\newcommand{\be}{\beta}
\newcommand{\sn}{\sin}
\newcommand{\cs}{\cos}
\newcommand{\hs}{\hspace{1cm}}
\newcommand{\lt}{\left}
\newcommand{\rt}{\right}
\newcommand{\mb}{\mbox}
\newcommand{\zg}{\lt(\frac{g}{\cos \theta_W}\rt)^2}
\newcommand{\lsim}%

\title{One-loop contributions of the super-partner particles to 
$e^-e^+ \rightarrow W^-W^+$ in the MSSM}
\author{Kaoru Hagiwara}

\address{Theory Group, KEK, Tsukuba, Ibaraki 305-0801, Japan\\
         E-mail: kaoru.hagiwara@kek.jp}

\author{Shinya Kanemura}
\address{Institut f\"{u}r Theoretishce Physik der Universit\"{a}t Karlsruhe, 
         D-76128 Karlsruhe, Germany\\
         E-mail: kanemu@physik.uni-karlsruhe.de}
 
\author{Yoshiaki Umeda}
\address{II Institut f\"ur Theoretishce Physik der Universit\"at Hamburg, 
         D-22761 Hamburg, Germany\\
         E-mail: umeda@mail.desy.de}  

\twocolumn[\maketitle\abstract{One-loop contributions 
of super-partner particles to $W$-pair production at $e^+e^-$ 
collision are discussed in the MSSM.  
To obtain trustworthy results we test our calculation using 
three methods: 
(1) sum rules among form factors   
    which result from the BRS invariance, 
(2) the decoupling theorem, 
(3) the high-energy stability. 
We examine the corrections taking into account constraints 
from the direct search experiments and the precision data.
The results for the sfermion contributions are presented. 
}]

\section{Introduction}%


We discuss the one-loop super-partner particle contributions 
to $e^-e^+ \to W^-W^+$ in the MSSM. 
The SM particles have their partners, such as sfermions 
and inos. We here concentrate on the sfermion one-loop effects\cite{sf}.
The sfermions include squarks and sleptons, 
whose mass matrices are expressed by 
{\small
\begin{eqnarray}
M^2_{\tilde{f}} = \!\!\! \left[ \begin{array}{cc}
\!\!\!
m_{Q,L}^2\!\! +\! m_Z^2 c_{2\be} 
   (T^3_{f_L}\!\!\! - \!\! \hat{s}^2 Q_{f}\!)\! +\!\! m_f^2 
   &\;   -\! m_f\! A^{\sf eff}_f \\ \\
\!\!\!\!\!\!\!\!\!\!\!\!\!\!\!\!\!\!\!\!\!\!\!\!\!\!\!\!\!\!\!\!\!\!\!\!
\!\!\!\!\!\!\!\!\!\!\!\!\!\!\!\!
-m_f A^{{\sf eff}\ast}_f &
\!\!\!\!\!\!\!\!\!\!\!\!\!\!\!\!\!\!\!\!\!\!\!\!\!\!\!\!\!\!\!\!\!\!\!\!
\!\!\!\!\!\!\!\!\!\!\!\!\!\! m^2_{U,D,E}  \!
+\! m_{\! Z}^{} c_{2\be} \hat{s}^2 Q_f \! +\! m_f^2 \!\!\!
\end{array} \right]\; . \nn 
\end{eqnarray}
}
The off-diagonal elements
$A_{D,E}^{\sf eff} = A_{D,E}+ \mu \tan\beta$ and 
$A_{U}^{\sf eff} = A_{U}+ \mu \cot\beta$ are multiplied by the fermion mass, 
so that the mixing  are important for stops.

\section{Calculation}

Helicity amplitudes for 
$e^-(k,\tau) e^+ (\overline{k},\overline{\tau})$ 
$\to W^-(p,\lambda) W^+(\overline{p},\overline{\lambda})$, 
where  
$k,\overline{k},p,\overline{p}$ are momenta,  
$\tau, \overline{\tau} (=-\tau), \lambda, \overline{\lambda}$ are helicities,   
 may be expressed by using 16 basis tensors as  
{\small 
\begin{eqnarray}
  {\cal M}_\tau^{\lambda\overline{\lambda}}\! = \!
\sum_{i=1}^{16} F_{i,\tau}(s,t)\, 
j_\mu 
\,T_i^{\mu\al\be}
\epsilon_\al^\ast(p,\lambda) 
\epsilon_\be^\ast(\overline{p},\overline{\lambda}). 
\end{eqnarray}}
The 16 form factors, $F_{i,\tau}$,  
include all information of the dynamics, 
while the other part are determined by the kinematics. 
For physical $W$-pair production, 9 basis tensors are enough. 
The rest are used for processes with  
unphysical (scalar) $W$ bosons, which are used for 
the test of $F_{i,\tau}$. 
For this test,  we also need to calculate   
$e^-(k,\tau) e^+ (\overline{k},\overline{\tau}) 
 \to W^-(p,\lambda) w^+(\overline{p})$   
($w^+$: the Nambu-Goldstone boson), 
whose amplitudes are decomposed as 
{\small 
\begin{eqnarray}
  {\cal M}_\tau^{\lambda}\! = \! i 
\sum_{i=1}^{4} H_{i,\tau}(s,t)\, 
j_\mu 
\,S_i^{\mu\al}
\epsilon_\al^\ast(p,\lambda), 
\end{eqnarray}
}
with four basis tensors and form factors $H_{i,\tau}$. 

We employ the $\overline{\rm MS}$ scheme, and 
we take $\hat{e}$, $\hat{g}$ and $M_W^{}$ as input SM parameters.   
The MSSM $\overline{\rm MS}$ couplings are determined by 
\begin{eqnarray}
  \fr{1}{\hat{e}_{\sf MSSM}^2 (\mu)}  
&=& 
  \fr{1}{\hat{e}_{\sf SM}^2 (\mu)} - \Delta \Pi_{T,\gamma}^{QQ}(0, \mu) ,  \\
  \fr{1}{\hat{g}_{\sf MSSM}^2 (\mu)}  
&=& 
  \fr{1}{\hat{g}_{\sf SM}^2 (\mu)} - \Delta \Pi_{T,\gamma}^{3Q}(0, \mu), 
\end{eqnarray}
where $\Delta \Pi_{T,\gamma}^{QQ}(0, \mu)$ and 
$\Delta \Pi_{T,\gamma}^{3Q}(0, \mu)$ are non-SM  
contributions to gauge-boson two-point functions.  
The SM $\overline{\rm MS}$ couplings  
are calculated by using the SM RGE's 
and experimental values for the effective charges\cite{hhkm}. 
$M_W=80.41$GeV is taken from the data.

\section{Tests}

One difficulty in loop-level calculations is to determine reliability 
of the results. This is especially so in our process in which a subtle 
gauge cancellation takes place among diagrams at each level of 
perturbation. Incomplete treatment for higher order terms can lead 
artificially large collections. In order to obtain solid results, we test 
our calculation by the following methods. 

\vspace*{-2mm}
\subsection{The BRS invariance}

Useful sum rules among form factors between 
the $W^-W^+$ and the $W^-w^+$ processes are induced from the 
BRS invariance;  
\vspace*{-2mm}
\begin{eqnarray}
  \sum_{j=1}^{16} \xi_{i j} F_{j, \tau}(s,t) 
= C_{\rm \sf mod} H_{i, \tau}(s,t),
\end{eqnarray}
where $\xi_{ij}$ are determined by the kinematic parameters 
and $C_{\rm \sf mod}$ differs from 1 at loop levels. 
We can use them 
to test  $F_{i,\tau}$. 
In Table 1, values for both sides in the sum rule 
are shown.  They coincide with each other. 

{\small
\begin{table}[t]
\begin{tabular}{l|l} 
\multicolumn{2}{l}{{\normalsize \sf Test by using the BRS sum rule} 
{\footnotesize($i= 1, \; \tau=-1$)}}
 \\ \hline
$\sqrt{s}$ & \sf  Left-hand-side $\;\;\;\;\;\;\xi_{1 j} F_j(s,t)$\\
           & \sf Right-hand-side $\;\;\;\;C_{\sf mod} H_i(s,t)$\\ \hline
%
%
\sf 
 200 GeV    &  \sf 
      $-$1.385496590672218 $\times {\sf 10}^{\sf -6}$ \\
           &  
\sf 
      $-$1.385496590672223 $\times {\sf 10}^{\sf -6}$ \\ \hline
%
%
%
%
\sf 
  1000GeV  & \sf 
$-$6.682526871892199 $\times {\sf 10}^{\sf -8}$ 
\\
           & \sf 
$-$6.682526871892053 $\times {\sf 10}^{\sf -8}$ 
\\ \hline
\end{tabular}
\caption{The test by the BRS sum rules.}
\end{table}
}

\vspace*{-2mm}
\subsection{The decoupling theorem}

The cross section should be of the SM prediction 
in the large sfermion-mass limit. We use this fact 
to test the overall renormalization factor 
which cannot be tested by the BRS sum rules.  
In Figure 1, for large-mass limit 
($1/M^2 \to 0$: $M$ is a scale of SUSY soft-breaking masses),  
the deviation from the SM prediction becomes zero for 
each case.

\begin{figure}
\epsfxsize160pt
\figurebox{160pt}{200pt}{decoup_B.epsi}
\caption{Test of decoupling. 
        Case A and B correspond to 
        non-mixing and stop-mixing cases.}
\label{fig:radk}
\epsfxsize160pt
\figurebox{160pt}{200pt}{m00_HBdy.epsi}
\caption{Test for high-energy stability. Real curve and dotted one 
corresponds to results from the full calculation and the analytic 
expression, respectively.}
\label{fig:radk}
\end{figure}

\vspace*{-4mm}
\subsection{High energy stability}

At high energies large gauge cancellation takes place, 
so it is important to see the high energy stability of the numerical
results. 
We calculate high-energy analytic expression for the amplitude.  
In Figure 2, the high-energy expression 
and the full calculation give same results in the high energy limit.

\section{Sfermion one-loop effects}


In Figure 3, the squark one-loop effects 
on the $00$ helicity amplitude for parameter sets in Table 2. 
The corrections to the SM prediction are negative and 
the behavior is rather simple. There is a
peak slightly after the squark-pair threshold. 
The corrections to the SM prediction are 
at most a few times 0.1\%.

{\small 
\begin{table}[t]
{
 \small
\begin{tabular}{l|rrr}
{\sf $\!\!\!\!$ First 2 generations} $\!\!\!\!\!\!$ 
& Case 1 & Case 2 & Case 3   \\ \hline
\multicolumn{4}{l}{Input parameters }           \\ \hline
$m_{\tilde{Q}}$= 
$m_{\tilde{U}}$=  
$m_{\tilde{D}}$ & 300 & 500 &1000 \\
$A_{\tilde{f}}^{eff}$&   0 &   0 &   0  \\ 
\hline  \hline
\end{tabular}}
\caption{Cases without mixing.}
\end{table}
}
\begin{figure}
\epsfxsize140pt
\figurebox{140pt}{180pt}{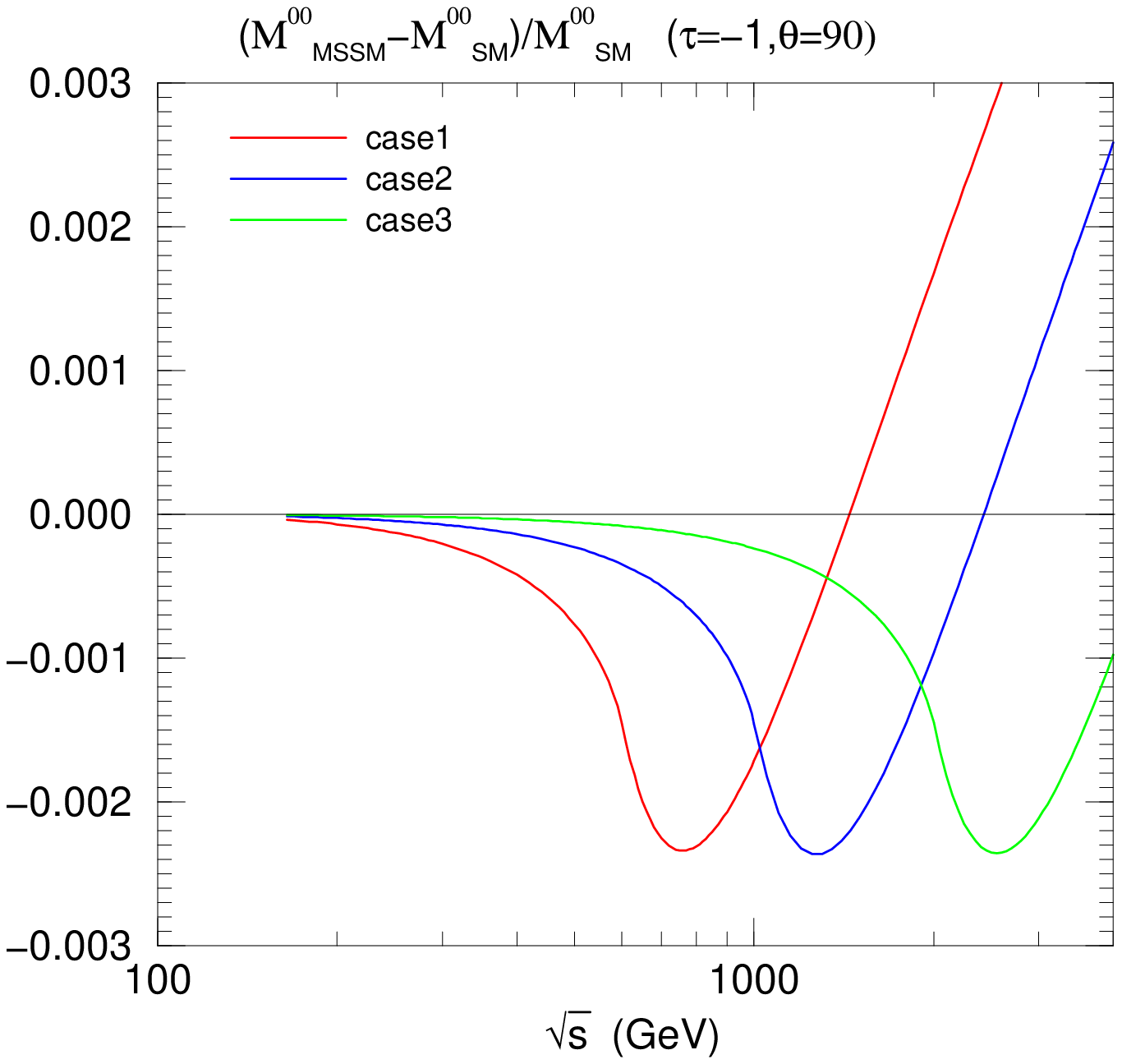}
\caption{
Squark effects of the first two generation. 
There is no mixing between $\tilde{f}_L$ and $\tilde{f}_R$}
\label{fig:radk}
\end{figure}


In Figure 4,  effects of the third generation squarks 
with large stop-mixing are shown.  The parameters defined in  
Table 3 are chosen so as to be the maximal mixing 
with mixing angle 45$^\circ$.
The corrections are positive.  Larger effects 
appear for larger $A^{\sf eff}_f$. 
It, however, turns out that such enhancement due to the 
mixing is strongly constrained by the precision data.  
In Figure 5, each case in Table 3 is plotted on the $S$-$T$ 
parameter plane\cite{sf}.  The cases for large corrections 
(case 2, case 3 in Table 3) 
stay outside the 99\% contour and thus they are excluded. 
After all only smaller corrections than a few times 0.1\% 
are allowed. 

\section{Conclusion}

The sfermion effects on this process is small.

\begin{table}[hph]
{\small
\begin{tabular}{l|rrr}
{\sf $\tilde{t}$-$\tilde{b}$ sector:}  & Case 1 & Case 2 & Case 3   \\ \hline
\multicolumn{4}{l}{Input parameters }           \\ \hline
$m_{\tilde{Q}}$= 
$m_{\tilde{U}}$=  
$m_{\tilde{D}}$ & 300 & 400 &500 \\
$A_{\tilde{f}}^{eff}$& 625 & 1025 & 1539 \\ \hline
\hline  
\multicolumn{4}{l}{Output parameters} \\ \hline
$m_{\tilde{t}_1}$ & 100 & 100 &  100 \\
$m_{\tilde{t}_2}$ & 478 & 607 & 741 \\
$\cos\theta_{\tilde{t}}$ &0.708&0.708&0.707 \\
\hline\hline 
\end{tabular}
\caption{Maximal stop-mixing cases.}
}
\end{table}
\vspace*{-0.5cm}
\begin{figure}[hph]
\epsfxsize140pt
\figurebox{140pt}{180pt}{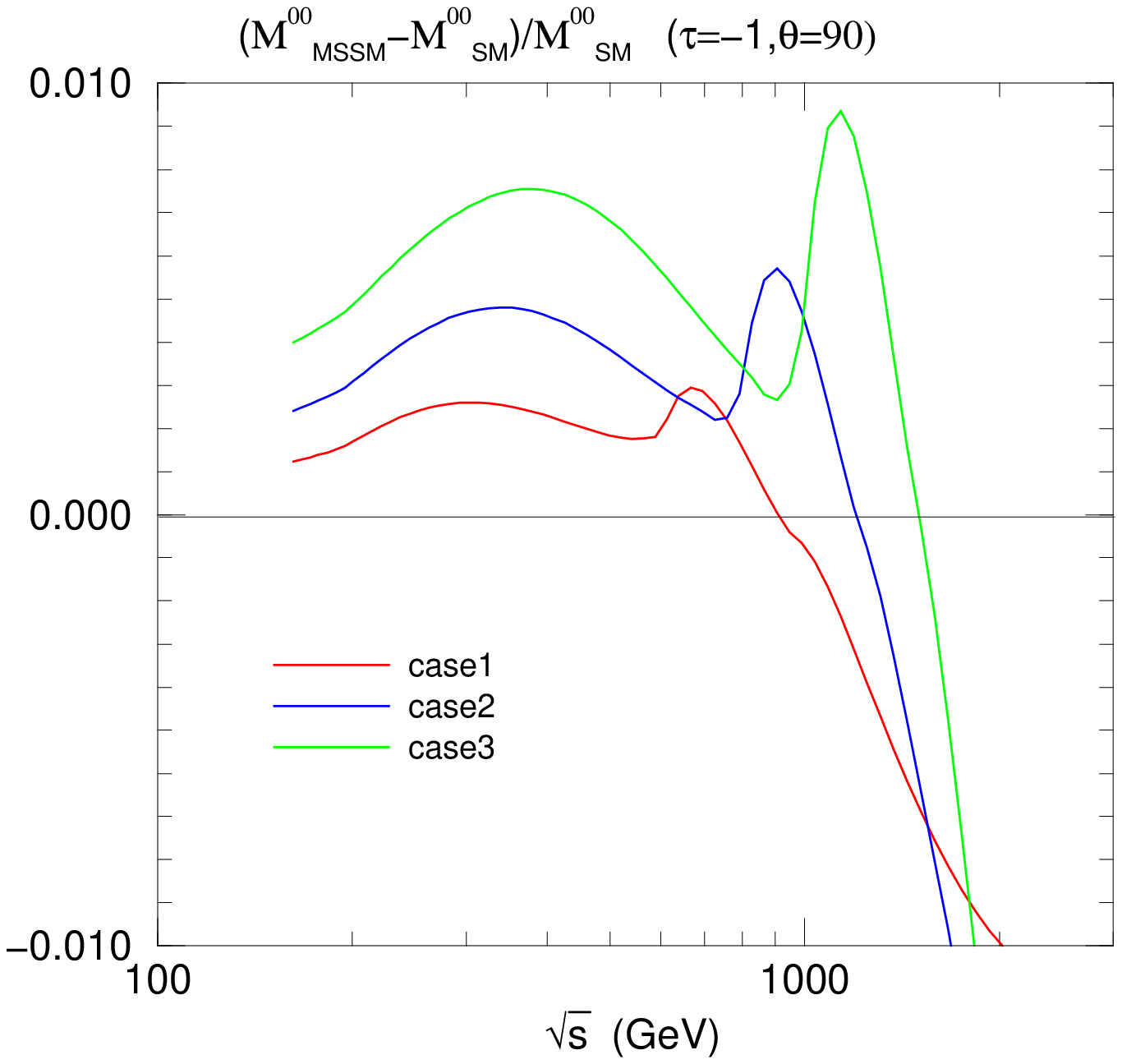}
\caption{The third-generation squark effects. 
        Maximal stop-mixing cases.}
\epsfxsize142pt
\figurebox{160pt}{180pt}{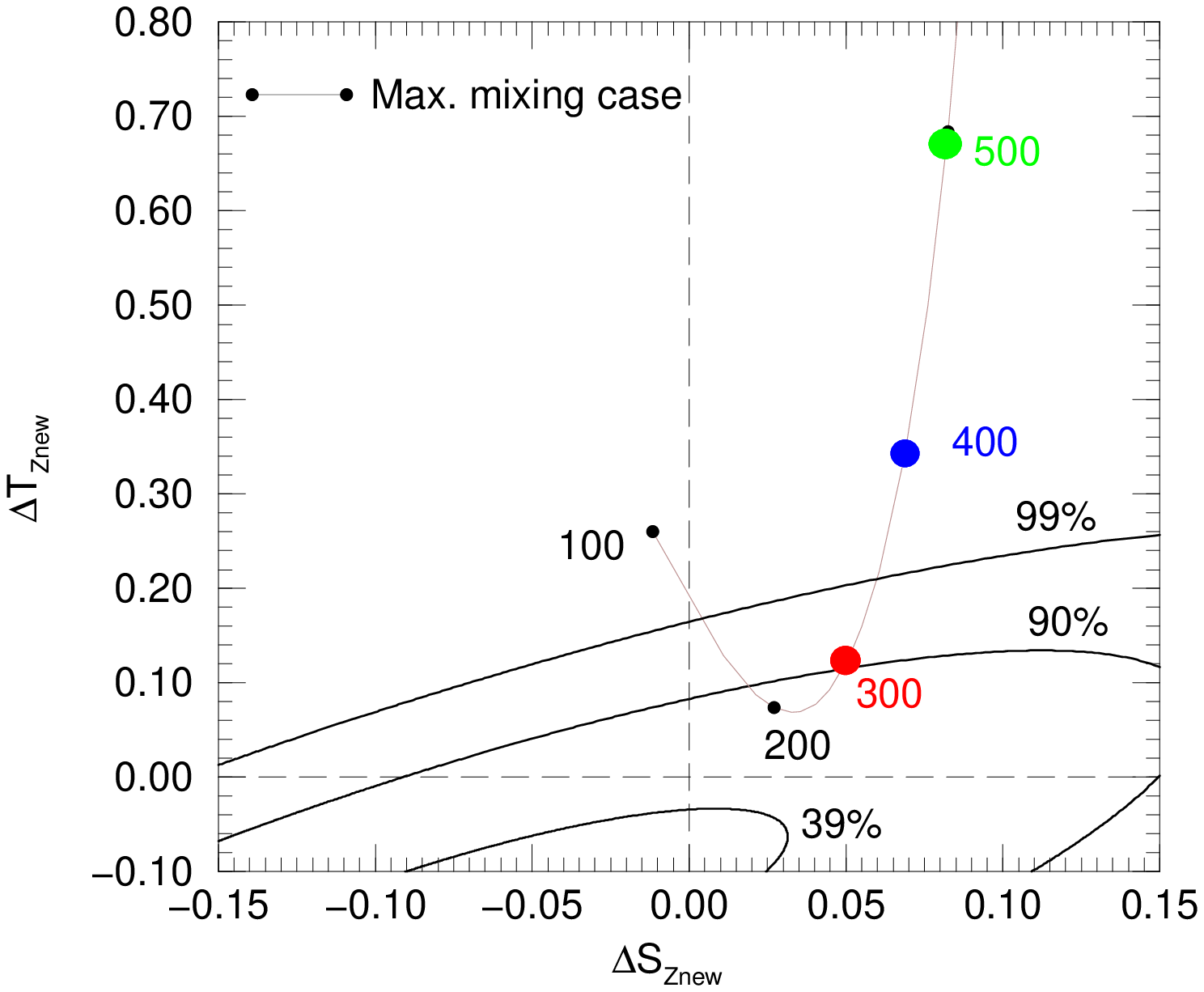}
\caption{The parameter sets defined in Table 3 on $S$-$T$ plane. The origin 
         indicates the SM prediction.}
\label{fig:radk}
\end{figure}


\begin{thebibliography}{99}

\bibitem{sf} S. Alam, K. Hagiwara, S. Kanemura, R. Szalapski, and Y. Umeda, 
             To appear in Phys. Rev. {\bf D}, ({\tt hep-ph/0002066}), and 
             references therein; 
             Nucl. Phys. {\bf B541} (1999) 50.  

     
\bibitem{hhkm} K. Hagiwara, D. Heidt, C.S. Kim and S. Matsumoto,
              Z. Phys. {\bf C64} (1994) 559.
   
\end{thebibliography}
\end{document}